\documentclass[a4paper,11pt]{article}
\pdfoutput=1 

\usepackage{jcappub} 
\usepackage[T1]{fontenc} 

\usepackage{graphicx}
\usepackage{dcolumn}
\usepackage{bm}
\usepackage[colorlinks=true]{hyperref}
\usepackage[mathlines]{lineno}
\usepackage{aas_macros}
\usepackage{color}
\usepackage{bbm}
\usepackage{multirow}
\usepackage{siunitx}
\usepackage{subcaption}
\usepackage{booktabs}
   
\newcommand{\dlum}{\ensuremath{d_\text{L}}}
\newcommand{\dcom}{\ensuremath{d_\text{M}}}
\newcommand{\dang}{\ensuremath{d_\text{A}}}
\newcommand{\dhub}{\ensuremath{d_\text{H}}}

\newcommand{\lcdm}{\ensuremath{{\Lambda\text{CDM}}}}

\newcommand{\Ok}{\ensuremath{\mathcal{O}_k}}
\newcommand{\Om}{\ensuremath{\mathcal{O}m}}

\newcommand{\Omo}{\ensuremath{\Omega_{\text{m},0}}}

\newcommand{\Oko}{\ensuremath{\Omega_{k,0}}}

\newcommand{\one}{\ensuremath{\mathbbm{1}}}
\newcommand{\Hord}{\ensuremath{H_0\rd}}
\newcommand{\vect}[1]{\ensuremath{\boldsymbol{#1}}}
\newcommand{\tens}[1]{\ensuremath{\bm{#1}}} 
\newcommand{\mat}[1]{\ensuremath{\bm{#1}}} 
\newcommand{\rd}{\ensuremath{r_\text{d}}} 
\newcommand{\diff} {\ensuremath{\mathrm{d}}} 
 
\newcommand{\deriv}[2]{\ensuremath{\frac{\diff {#1}}{\diff {#2}}}}

\newcommand{\DD}{\ensuremath{\mathcal{D}}}

\usepackage{scalerel}
\usepackage{tikz}
\usetikzlibrary{svg.path}

\definecolor{orcidlogocol}{HTML}{A6CE39}
\tikzset{
  orcidlogo/.pic={
    \fill[orcidlogocol] svg{M256,128c0,70.7-57.3,128-128,128C57.3,256,0,198.7,0,128C0,57.3,57.3,0,128,0C198.7,0,256,57.3,256,128z};
    \fill[white] svg{M86.3,186.2H70.9V79.1h15.4v48.4V186.2z}
                 svg{M108.9,79.1h41.6c39.6,0,57,28.3,57,53.6c0,27.5-21.5,53.6-56.8,53.6h-41.8V79.1z M124.3,172.4h24.5c34.9,0,42.9-26.5,42.9-39.7c0-21.5-13.7-39.7-43.7-39.7h-23.7V172.4z}
                 svg{M88.7,56.8c0,5.5-4.5,10.1-10.1,10.1c-5.6,0-10.1-4.6-10.1-10.1c0-5.6,4.5-10.1,10.1-10.1C84.2,46.7,88.7,51.3,88.7,56.8z};
  }
}

\newcommand\orcid[1]{\href{https://orcid.org/#1}{\mbox{\scalerel*{
\begin{tikzpicture}[yscale=-1,transform shape]
\pic{orcidlogo};
\end{tikzpicture}
}{|}}}}

\definecolor{deepmagenta}{rgb}{0.8, 0.0, 0.8}
\definecolor{ballblue}{rgb}{0.13, 0.67, 0.8}

\definecolor{RedWine}{rgb}{0.743,0,0}

\title{Litmus tests of the flat \lcdm\ model and model-independent measurement of \Hord\ 
with LSST and DESI
}

\author[a]{Benjamin L'Huillier\orcid{0000-0003-2934-6243},}
\author[b,c]{Ayan Mitra\orcid{0000-0002-9436-8871},}
\author[d,e]{Arman Shafieloo\orcid{0000-0001-6815-0337},}
\author[f]{Ryan E Keeley\orcid{0000-0002-0862-8789}}
\author[g]{and Hanwool Koo\orcid{0000-0003-0268-4488}}

\affiliation[a]{Department of Physics and Astronomy, Sejong University, 05006 Seoul, Korea}
\affiliation[b]{Center for AstroPhysical Surveys, National Center for Supercomputing Applications, University of Illinois Urbana-Champaign, Urbana, IL, 61801, USA}
\affiliation[c]{Department of Astronomy, University of Illinois at Urbana-Champaign, Urbana, IL 61801, USA}
\affiliation[d]{Korea Astronomy and Space Science Institute, 
Daejeon 34055, Korea}
\affiliation[e]{University of Science and Technology, 
Daejeon 34113, Korea}
\affiliation[f]{University of California, Merced, 5200 N Lake Road, Merced, CA 95341, USA }
\affiliation[g]{Department of Physics and Astronomy, University of the Western Cape, Robert Sobukwe Road,
Bellville, Cape Town, 7535, South Africa}

\emailAdd{benjamin@sejong.ac.kr}
\emailAdd{ayan@illinois.edu}
\emailAdd{shafieloo@kasi.re.kr}
\emailAdd{rkeeley@ucmerced.edu}
\emailAdd{hanwool.koo90@gmail.com}

\arxivnumber{2407.07847} 
\definecolor{owngreen}{rgb}{0.0, 0.5, 0.0}

\abstract{In this analysis we apply a model-independent framework to test the flat $\Lambda$CDM cosmology using simulated SNIa data from the upcoming Legacy Survey of Space and Time (LSST)
and combined with simulated Dark Energy Spectroscopic Instrument (DESI) five-years Baryon Acoustic Oscillations (BAO) data. We adopt an iterative smoothing technique to reconstruct the expansion history from SNIa data, which, when combined with  BAO measurements, facilitates a comprehensive test of the Universe’s curvature and the nature of dark energy. The analysis is conducted under four different mock true cosmologies: 
Two curvatures ($\Oko=0$ and 0.1) and two models of dark energy: a cosmological constant $\Lambda$ and the phenomenologically emergent dark energy. 
We forecast that our reconstruction technique can constrain cosmological parameters, such as the curvature ($\Oko$) and  $c/H_0\rd$, with 
spread due to the SNIa uncertainties up to $\pm 4\%$ and $\pm 0.1$ respectively, 
without assuming any form of dark energy.
}

\begin{document}

\maketitle 

\flushbottom

\section{Introduction} 

The concordance cosmological \lcdm\ model relies on a few assumptions. 
Among them, homogeneity and isotropy lead to the Friedman-Robertson-Lemaître-Walker (FLRW) metric. 
The discovery of cosmic acceleration led to the establishment of the concordance \lcdm\ model. 
In this model,  the energy budget of the Universe today is dominated by a cosmological constant $\Lambda$, and matter by cold dark matter (CDM). 
The most stringent constraints on the model have been obtained by cosmic microwave background (CMB) measurements from the Planck satellite \cite{Pl18VI}, baryon acoustic oscillations (BAO) \cite{2005ApJ...633..560E,
2021PhRvD.103h3533A,
2024arXiv240403001D,
2024arXiv240403002D,
2024arXiv240403000D} and Type Ia Supernovae (SNIa) \cite{perl,riess_1998,Betoule2014,Pantheon,pantheon_new}.

The rise of tensions, in particular the Hubble tension, and the unknown nature of the main components of the model are strong hints for cosmologists to search for evidence for beyond-\lcdm\ physics \cite{Aluri:2022hzs, Perivolaropoulos:2021jda, Kamionkowski:2022pkx, DiValentino:2021izs,Krishnan:2021dyb,Abdalla:2022yfr,Krishnan:2021jmh,Smith:2022hwi,SolaPeracaula:2022hpd}.
While the simplest explanation for the late-time cosmic acceleration is dark energy in the form of a cosmological constant, a dynamical DE is not excluded \cite{2017NatAs...1..627Z}. 
Recent BAO data from the Dark Energy Spectroscopic Instrument (DESI) showed evidence for a phantom dark energy or a dynamical dark energy \cite{2024arXiv240403002D}. 
This was further confirmed by studying physics-driven modifications of \lcdm\ \cite{2024arXiv240513588L}
as well as data-driven reconstructions of the expansion history \cite{2024arXiv240504216C}. 

Another fundamental question is the curvature of the Universe, summarized in the current curvature density parameter \Oko.
The Planck 2018 (P18) data favor a negative curvature $-0.095<\Oko<-0.007$ at 99\% \cite{Pl18VI,2020NatAs...4..196D,2021PhRvD.103d1301H}, 
However, combining P18  with external data yields results consistent with a flat universe \cite{2021PDU....3300851V,2021ApJ...908...84V}.
Interestingly, DESI DR1 finds a preference for a small positive curvature, which disappears when combined with Planck.  

Most of these constraints however are direct fits to the data assuming a particular model, typically the \lcdm\ model or an extension. 
While this leverages the power of Bayesian statistics to provide tight constraints on the model parameters, it is important to also test the validity of the underlying hypotheses. 
As opposed to model-fitting, model-independent approaches provide a healthy check on the validity of a given model.
Due to their extra flexibility compared to model-fitting approaches, they may also reveal the presence of unexpected features in the data and may be used to look for unaccounted for systematics~\cite{2021AJ....161..151K,2023JCAP...02..014H,2023PhRvL.131k1002K}.

In this work, we used a model-independent approach, namely, the iterative smoothing method, to reconstruct the expansion history from SNIa and combine it with distance measurements from BAO to perform litmus tests of the pillars of the \lcdm\ model, the FLRW metric, the flatness of the Universe, and the cosmological constant, as well as to constrain $c/\Hord$.
Specifically, our goal is twofolds: we aim to quantify the constraining power of the combination of LSST SNIa and DESI Y5 BAO to (1) test the FLRW metric, and (2) to measure the curvature parameter and \Hord\ independently of the nature of DE.

\section{Theory}

In an FLRW universe, when radiation is negligible, the expansion history reads
\begin{subequations}
\begin{align}
    h^2(z) & = \frac{H^2(z)}{H_0^2} = \Omo(1+z)^3 + \Oko(1+z)^2 +(1-\Omo - \Oko) f_\text{DE}(z), \\
    \intertext{where}
    f_\text{DE}(z) &=  \exp\left(3\int_0^z \frac{1+w(x)}{1+x}\diff x\right)
\end{align}
\end{subequations}
describes the time evolution of the dark energy density, 
 \Omo\ and \Oko\ are the matter and curvature density parameters today respectively, and $w = p/\rho$ is the equation of state of dark energy. 
 For the cosmological constant $\Lambda$, $w=-1$ and $f_\text{DE}(z) \equiv 1$.

The dimensionless comoving distance is defined by
\begin{align}
\label{eq:D}
    \DD(z) & = \frac 1 {\sqrt{-\Oko}} \sin\left[ \sqrt{-\Oko} \int_0^z \frac{\diff x}{h(x)}\right]  
\end{align}
for all signs of \Oko. In particular, for $\Oko=0$, eq.~\eqref{eq:D} simplifies to 
\begin{align}
    \DD(z) & = \int_0^z \frac {\diff x}{h(x)}.\label{eq:Dflat}
\end{align}

The angular diameter and luminosity distances are related to the comoving distance by
\begin{align}
    (1+z) \dang(z) &  = \frac{\dlum(z)}{1+z}= \dcom(z) = \frac c {H_0} \DD(z).
\end{align}

The distance modulus from supernovae is 
\begin{align}
    \mu(z) & = 5\log_{10} \frac{\dlum}{\SI 1 {Mpc}} + 25. 
\end{align}

BAO, on the other hand, measure both the distances and expansion. 
Strictly speaking, the transverse and radial modes of the BAO give 
\begin{subequations}
\begin{align}
    \frac{\dang(z)}{\rd} & = \frac c {\Hord} \frac   {\DD(z)}{1+z} 
    \mbox{, and}\\
    \frac{c}{H(z)\rd} & = \frac c{\Hord} \frac 1 {h(z)}
    \mbox{, where}\\
    \frac \Hord c & = \frac 1 {\sqrt 3}  \int_0^{\tfrac 1 {1+z_\mathrm{d}}} \frac{\diff a}{a^2h(a) \sqrt{1+\frac{3\omega_b}{4\omega_\gamma}a}} 
\end{align}
\end{subequations}
and \rd\ is the sound horizon at the drag epoch $z_\mathrm{d}$.
$H_0\rd$ is thus an interesting parameter combining early- and late-time physics, which may play an important role in the $H_0$ tension~\cite{2017JCAP...01..015L,
2018PhRvD..98h3526S,
Staicova:2021ntm,
Benisty:2024lmj}.

\section{Data}
\label{sec:data}

In this analysis, we used simulated $3$-year LSST\footnote{\url{www.lsst.org}} SNIa data and the BAO forecast for DESI from \cite{2016arXiv161100036D} to constrain cosmology.

\subsection{Fiducial Models}
\label{sec:fid}
To show the potential of our model-independent approach, we simulate data for
four different fiducial cosmologies. All fiducial cosmologies share the same $(\Omo, h) = (0.315,0.7)$, which is the cosmology used in \cite{Mitra:2022ykq}. 
\begin{itemize}
\item Flat-$\Lambda$: a flat-\lcdm\ universe, with  
a constant equation of state $w=-1$.
\item $k$-$\Lambda$: a curved \lcdm\ universe, with
$\Oko = 0.1$.
While such a large curvature is consistent with Pantheon+ or DESI DR1 data, it is excluded by Planck 2018 data. 
However, it is nonetheless interesting to see the effect of curvature and to apply our curvature test to a curved universe.
\item Flat-PEDE: a flat universe with dynamical dark energy. We chose the phenomenologically emergent dark energy (PEDE) model \cite{2019ApJ...883L...3L}, which has no extra degree of freedom relative to \lcdm. 
The dark energy density evolves as 
\begin{subequations}
\begin{align}
    f_\text{PEDE}(z) & = 1-\tanh\left(\log_{10}(1+z)\right) \label{eq:fpede}
    \intertext{and the equation of state}
    w_\text{PEDE}(z) & = \frac 1 {3\ln 10} \left(1+\tanh\left[\log_{10}(1+z)\right]\right)-1 \label{eq:wpede}
\end{align} 
\end{subequations} 
\item  $k$-PEDE: a curved universe with $\Oko=0.1$ and PEDE.
\end{itemize}

We note that the SNIa mock data and covariance matrix were initially generated for the flat \lcdm\ case.
We extracted the noise $\mu-\mu_\text{fid}$ and applied it to the other three fiducial cosmologies.  
Although for a proper forecast, one should apply the pipeline from \cite{Mitra:2022ykq} to the other three fiducial cosmologies, we assume here that the resulting noise and covariance should be affected minimally by the change of fiducial cosmology. 
On the other hand, the mock BAO data are all generated with a Gaussian white noise as detailed in Section~\ref{sec:mockbao}.

Table~\ref{tab:fidcosmo} summarizes our {four}  fiducial cosmologies. 

\begin{table}[]
    \centering
    \begin{tabular}{clcrr}
    \toprule
       Fiducial Cosmology   &     
       \Oko & $w$ & $\chi^2_{\lcdm}$ & $\Delta\chi^2_\mathrm{fid}$\\
       \midrule
        \lcdm\ &  
        0 &-1 & 7.48 & 1.27\\
        $k$-\lcdm &
        0.1 & -1 & 11.68 & -2.93\\
        PEDE &  
        0 & Eq.~\eqref{eq:wpede} & 8.21 & 0.54\\
        $k$-PEDE  &  
        0.1 & Eq.~\eqref{eq:wpede} & 7.55 & 1.20\\
        \bottomrule
    \end{tabular}
    \caption{Fiducial cosmologies used in this work. 
    The {fourth column shows the \lcdm\ $\chi^2_{lcdm}$ to the data,  used as a reference throughout this work, and the 
    last column shows the $\Delta\chi^2$ of the fiducial model with respect to the best-fit \lcdm.}}
    \label{tab:fidcosmo}
\end{table}

\subsection{LSST SNIa Simulations}
\subsubsection{The Rubin Observatory Legacy Survey of Space and Time (LSST)}
\label{sec:lsst}
The forthcoming {\tt LSST} survey, scheduled to commence in 2025, represents the most comprehensive optical survey planned to date, marking a significant milestone for its generation. 
This survey will run for a decade and will extensively explore the southern hemisphere's sky through a large, ground-based, wide-field observatory utilising six optical passband filters. 
This combination is designed to achieve an unparalleled balance of depth, coverage area, and observational frequency. 
Over the course of the survey, the LSST is expected to catalogue millions of supernovae~\cite{LSST:2008ijt}. Operating from the Vera C. Rubin Observatory, the Simonyi Survey Telescope features an $8.4$ m mirror (with an effective aperture of $6.7$ m) and is equipped with a state-of-the-art $3200$-megapixel camera, providing a $9.6$~deg$^2$ field of view. Approximately $90\%$ of the telescope's time will be dedicated to a deep-wide-fast survey mode, systematically covering an area of $18,000$ square degrees roughly $800$ times across all bands over a decade, resulting in a co-added map with a depth of $r-27.5$ (called as the Wide Fast Deep (WFD) observing strategy in the LSST). 
This extensive data collection is projected to amass around $32$ trillion observations of $20$ billion galaxies and an equivalent number of stars, primarily supporting the survey's main scientific objectives~\citep{LSST:2008ijt}. The remaining $10\%$ of the time will be reserved for specific initiatives, such as the Very Deep and Very Fast time-domain surveys,\footnote{\href{https://www.lsst.org/scientists/survey-design/ddf}{Deep Drilling Fields (DDF)}} which are still in the planning stages.

\subsubsection{Mock Data Generation}
\label{sec:snia}
This study employs simulated SNIa data to examine dark energy results from the forthcoming LSST observations, as detailed in \cite{Mitra:2022ykq}. The supernova data are generated through the LSST Dark Energy Science Collaboration (DESC) Time Domain (TD) pipeline and the {\tt SNANA} software \cite{snana}. In this analysis, our SN data refers to using the corresponding Hubble diagram, and the covariance matrix provided by  \cite{Mitra:2022ykq}. The Hubble diagram consists of the redshift ($z$) of measurements, the corresponding luminosity distance ($D_L$), and their statistical plus systematic errors combined ($\sigma (D_L)$).
The procedure encompasses four primary stages, depicted in Figure 4 of \cite{Mitra:2022ykq}.

The simulations utilise the flat \lcdm\ model from \S~\ref{sec:fid}. 
In particular, the Hubble constant, $H_0$, is set at $70.0~ \text{km s}^{-1} \text{Mpc}^{-1}$, aligned with the SALT2 model training.  
The SALT2 light curve model is employed to derive observer frame magnitudes. The simulation noise is calculated using the following equation:
\small
\begin{equation}
    \sigma_{\text{SIM}}^2 = \left[ F^2 + (A\cdot b)^2 + (F\cdot \sigma_{\text{ZPT}})^2 + \sigma_0\cdot 10^{0.4\cdot ZPT_{\text{pe}}} + \sigma_{\text{host}}^2 \right]S^2_{\text{SNR}}.
\end{equation}
\normalsize
where $F$ represents the simulated flux in photoelectrons, $A$ denotes the noise equivalent area, given by $A = \left[2\pi\int \text{PSF}^2 (r,\theta)r \mathop{}\!\mathrm{d} r\right]^{-1}$, with PSF standing for the Point Spread Function, and $b$ represents the background per unit area (inclusive of sky, CCD readouts, and dark current). The scale factor $S_{\text{SNR}}$ is empirically determined based on the signal-to-noise ratio. The terms represented by $\sigma$ denote uncertainties related to zero point, flux calibration, and the underlying host galaxy, empirically fitted to match simulated uncertainties with those derived from the survey designs.
 
The LSST TD pipeline involves SN brightness standardisation via a Light Curve (LC) fit stage, simulations for bias correction, and a BEAMS with Bias Corrections (BBC) stage for Hubble diagram production. This pipeline yields a redshift-binned Hubble diagram and its associated covariance matrix, which are the data products employed for cosmological fitting.

The SNIa dataset utilised comprises both spectroscopically ($z_{\rm spec}$) and photometrically ($z_{\rm phot}$) identified SNIa candidates. Based on {\tt PLAsTiCC} \cite{plasticc_announce}, the ``$z_{\rm spec}$'' sample includes two sets of events with spectroscopic redshifts featuring an accuracy of $\sigma_{z} \sim 10^{-5}$. The first subset consists of spectroscopically confirmed events with accurate redshift predictions by the 4MOST spectrograph \cite{4MOST2}, under construction by the European Southern Observatory and expected to be operational in 2024, situated at a latitude similar to that of the Rubin Observatory in Chile. The second subset includes photometrically identified events with accurate host galaxy redshifts determined by 4MOST, with this subset being approximately $60\%$ larger than the first. For the photometric sample, host galaxy photometric redshifts were employed as priors (adapted from \cite{Kessler2010}). The photometric redshift and rms uncertainty derive from \cite{Graham2018_photoz}. The entire simulation was rerun based on the {\tt PLAsTiCC} DDF data\footnote{The LSST utilises different observing strategies, namely the deep field and the wide field, termed the DDF (Deep Drilling Field) and the WFD (Wide Fast Deep), respectively.}. Additional low redshift spectroscopic data was sourced from the DC2 analysis, simulated with WFD cadence. \cite{Mitra:2022ykq} restricted simulations to SNIa, omitting contamination (\textit{e.g.} core collapse, peculiar SNe, \textit{etc}). The covariance matrix corresponds to the combined statistical and systematic errors, where the latter encompasses all individual systematics, detailed in Table 3 of \cite{Mitra:2022ykq}. The Hubble diagram is segmented into $14$ redshift bins, containing a total of $5809$ SNIa candidates, both spectroscopic and photometric. 

It is noteworthy that \cite{Mitra:2022ykq} introduce a dataset with characteristics akin to those projected in the LSST science roadmap for 1 and 10 years of SNIa cosmology analysis, as described by \cite{LSSTDarkEnergyScience:2018jkl}. However, there are significant differences. While the analysis by \cite{LSSTDarkEnergyScience:2018jkl} relies solely on SNIa with spectroscopically confirmed host redshifts, \cite{Mitra:2022ykq} broadens the scope by including a comprehensive end-to-end analysis that incorporates both spectroscopic and photometric redshifts of host galaxies. In the forthcoming LSST era, it is anticipated that the volume of photometrically observed Type Ia supernovae will vastly outnumber the spectroscopic candidates. The methodology presented in \cite{Mitra:2022ykq} therefore enables the use of these photometric candidates in the cosmology analysis and hence in construction of a denser Hubble diagram with higher redshift reaches. As a result, this  presents a critical opportunity to leverage photometric supernovae in cosmological analyses, and potentially amplifying the dark energy constraints significantly.

We would also like to mention that the SN analysis done using the LSST Time-Domain pipeline used a Gaussian SN likelihood, but the method presented in our analysis here, can as well incorporate asymmetric and non-Gaussian errors. Therefore our methodology {will still apply}, even if we had a different covariance matrix and likelihood for our data (including non-gaussian likelihoods).
{%
The fourth column of Table~\ref{tab:fidcosmo} shows the \lcdm\ $\chi^2$ to the SNIa data for each fiducial cosmology. As expected, the flat-$\Lambda$ cosmology has the lowest $\chi^2_{\lcdm}$. 
The last column shows $\Delta\chi^2=\chi^2_\text{fid}-\chi^2_{\lcdm}$, the difference between the $\chi^2$ of the fiducial cosmology to that of the \lcdm\ best-fit. 
It is worth noting that, since we use the same noise for all cosmologies, $\chi^2_\text{fid}$ is the same for all four cosmologies, in this case 8.75. 
Therefore, $\Delta\chi^2_\text{fid}$ is larger for the flat-$\Lambda$ cosmology. 
}

\subsection{Mock BAO data from DESI 5-year}
\label{sec:mockbao}

The Dark Energy Spectroscopic Instrument (DESI) is a stage-IV experiment aiming to constrain dark energy by mapping the Universe in three dimensions \cite{2024arXiv240403000D}. 
In particular, DESI is measuring the BAO, a probe of the expansion history from large-scale structures. 

We generated DESI 5-year mock BAO, assuming a Gaussian white noise where the  errors on \dang\ and $H$ are taken from \cite{2016arXiv161100036D}, assuming a correlation coefficient of 0.4 between $\dang$ and $H$, and then propagated to errors on $\dang$ and \dhub\ via the usual error propagation formula for correlated variables.

We also assume $H_0 = \SI{70}{km.s^{-1}.Mpc^{-1}}$, and 
$\rd^\text{fid} = \SI{144.0533}{Mpc}$, leading to a fiducial value of $\Hord=\SI{10083.731}{km.s^{-1}}=c/29.73$).
This fiducial value of $\rd$ can be obtained for the \lcdm\ fiducial cosmology with $\omega_b = 0.02237$.
While the value of $\rd$  would be calculated to be different for the other three fiducial cosmologies, we chose fix the value of \rd\ to the same value for the four fiducial cosmologies.  
While such a choice of $H_0$  is not currently favored by CMB experiments, the exact choice of cosmological parameters is not important for the purpose of this study since we are focusing on model independent reconstructions and so our results do not depend on different fiducial models.

\section{Method}

Several litmus tests have been put forward to test various aspects of the \lcdm\ model. 
In particular, the \Ok\ diagnostic \cite{2008PhRvL.101a1301C,2010PhRvD..81h3537S,2017JCAP...01..015L} is defined as 
\begin{subequations}
\begin{align}
\label{eq:Ok}
    \Ok(z) & = \frac{\Theta^2(z)-1}{\DD^2(z)},\intertext{where} 
    \Theta(z) & = h(z) \DD'(z), \label{eq:Theta}
\end{align}
\end{subequations}
{where $'$ denotes the derivative with respect to redshift.}
In an FLRW universe, $\Ok(z) \equiv \Oko$, and $\Theta(z) = \sqrt{1+\Oko\DD^2}$, and 
in a flat-FLRW universe, $\Theta(z) \equiv 1$. It is important to realize that these {equalities} only assume the FLRW metric, and in particular, are independent of the dark energy evolution. 

The $\Om$ diagnostic, introduced by \cite{2008PhRvD..78j3502S}, is defined as 
\begin{align}
    \Om(z) & = \frac{h^2(z)-1}{(1+z)^3-1} = \Omo \label{eq:Om}
\end{align}
in a flat-\lcdm\ universe.
These two diagnostics are litmus tests respectively of the FLRW metric (and its curvature) and of the flat-\lcdm\ model. 
Departure from $\Om=$ constant, or from $\Ok=$ constant are signs of departure from flat-\lcdm\ and FLRW respectively. 

Another widely used quantity is the deceleration parameter 
\begin{align}
    q & = -\frac 1 H \frac{\ddot a}{a} = -(1+z) \frac{h'}h-1,
\end{align}
{where $\dot{}$ denotes the derivative with respect to time $t$.}
$q>0$ indicate a decelerating Universe, while $q<0$ for an accelerating universe.

\subsection{Curvature test}

The transverse and radial BAO modes respectively give us ${\dang}/{\rd} = \dcom/((1+z)\rd)$ 
 and $ \dhub/\rd=c/{(H\rd)}$, 
while the supernovae provide $\DD$ and $\DD'$, which are independent of both $H_0$ and $\rd$. 
{Intuitively, the SNIa constrain (the log of) $\DD$, and we can obtain $\DD'$ by calculating the smooth derivative of \DD (see Section~\ref{sec:smooth}).}
We can therefore combine these two measurements to estimate \Hord\ in two 
independent ways: 
\begin{subequations}
\begin{align}
    \frac c{\Hord} & =  \frac 1 {\DD(z)} \frac {\dcom(z)}{\rd}\mbox{, and} \label{eq:HordD}\\
      \frac c{\Hord} & = h(z) \frac c{H\rd}  = \frac 1{\DD'(z)} \frac{\dhub(z)}{\rd}, \label{eq:HordH}
\end{align}
\end{subequations}
where the last equality in eq.~\eqref{eq:HordH} holds only for a flat-FLRW universe. 
Therefore, we can combine the BAO observables with $\DD$ and $\DD'$ measured by SNIa to constrain \Hord. 
One way to do this is to reconstruct a smooth comoving distance $\DD(z)$, its derivative $\DD'(z)$, and expansion history $h(z)$ from SNIa at any arbitrary redshift $z$, and evaluate those curves at the BAO redshifts $z_\mathrm{BAO}$. 
To do the reconstructions, we apply the iterative smoothing method to simulated LSST SNIa as described in Section~\ref{sec:smooth}.
We then simply divide the central value of the BAO measurements $\dcom/r_{\rm d}$ and $\dhub/r_{\rm d}$ by the value of the reconstructed ${\DD }$ and $\DD'$ evaluated at the BAO redshifts 
and then propagate the BAO error. 

We can then calculate 
\begin{align}
    \Theta(z) &  = \frac{H(z) \rd}{\Hord}  \DD' (z)
    =\frac{\dcom(z)/\rd}{\dhub(z)/\rd} \frac {\DD'(z)}{\DD(z)}
\end{align}
and \Ok\ only with the BAO and the smooth reconstructions from the SNIa data. 
Similarly to $H_0r_{\rm d}$, we evaluate $\Theta(z)$ by multiplying the central values of the BAO data ($\dhub/\rd$ and ${\dcom}/{\rd}$) with the smooth functions reconstructed from the SNIa 
($\DD'/\DD$), evaluated at the BAO redshifts $z_{\rm BAO}$, and propagating the error by taking the covariance between $\dcom$ and $\dhub$ as detailed in Appendix~\ref{sec:errorprop}.

\begin{figure}
    \centering
    \includegraphics[width=\textwidth]{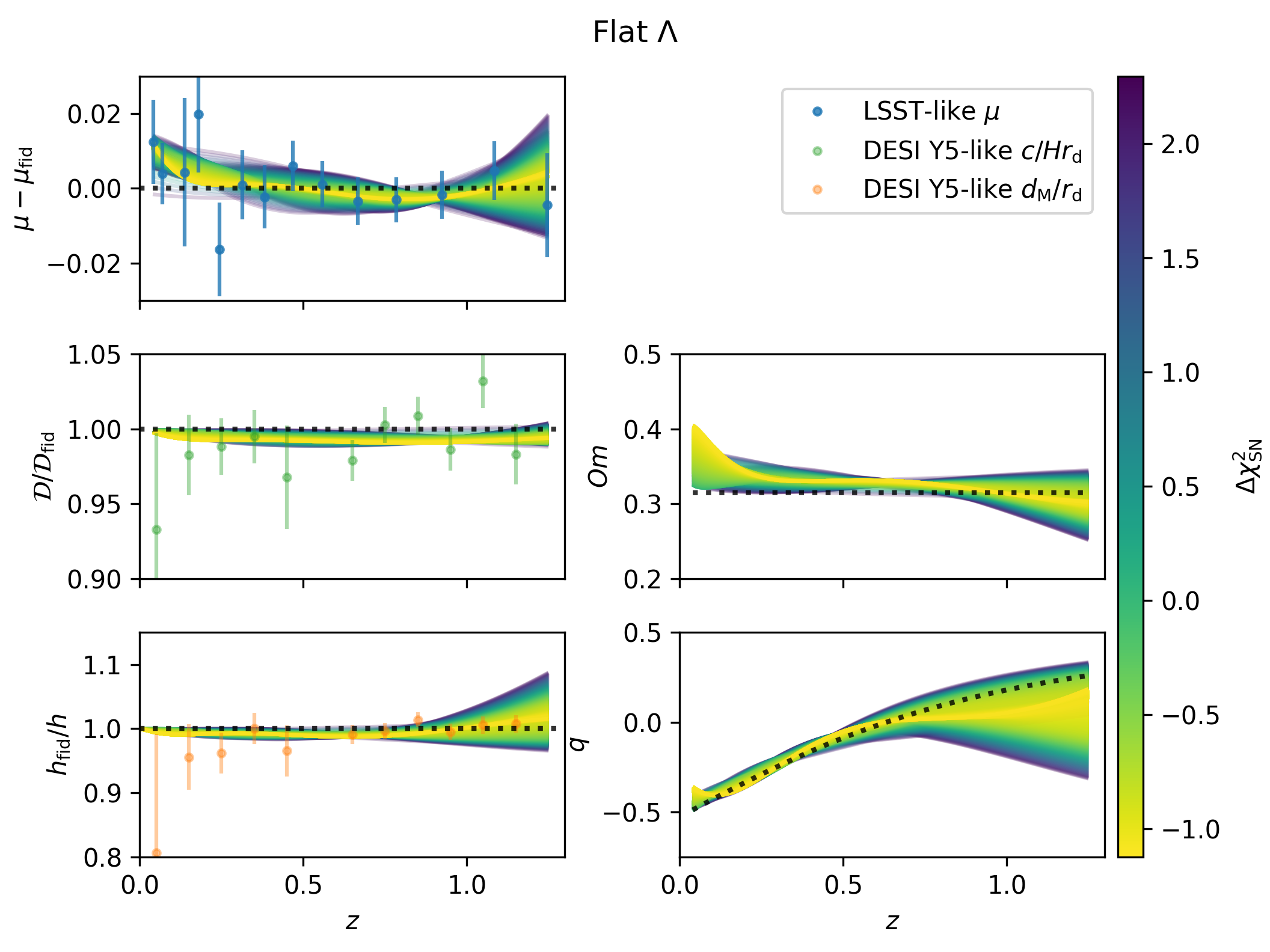}
    \caption{
    \label{fig:smoothing_lcdm} 
      Results of the SNIa smoothing colour-coded by $\Delta\chi^2_{\text{SN},n} = \chi^2_{\text{SN},n}-\chi^2_\text{SN, \lcdm}$ for the flat-\lcdm\ fiducial cosmology. 
      The top-left panel shows the smooth residuals. 
      The middle-left panel shows $\DD/\DD_\text{fid}$, the bottom-left panel shows  $h_\text{fid}/h$,  the middle-right panel shows $\Om$, and the bottom-right panel $q$.
      The middle- and bottom-left panels also show the DESI-like BAO mock data normalized by the fiducial cosmology. 
    }
\end{figure}

\begin{figure}
    \centering
    \includegraphics[width=\textwidth]{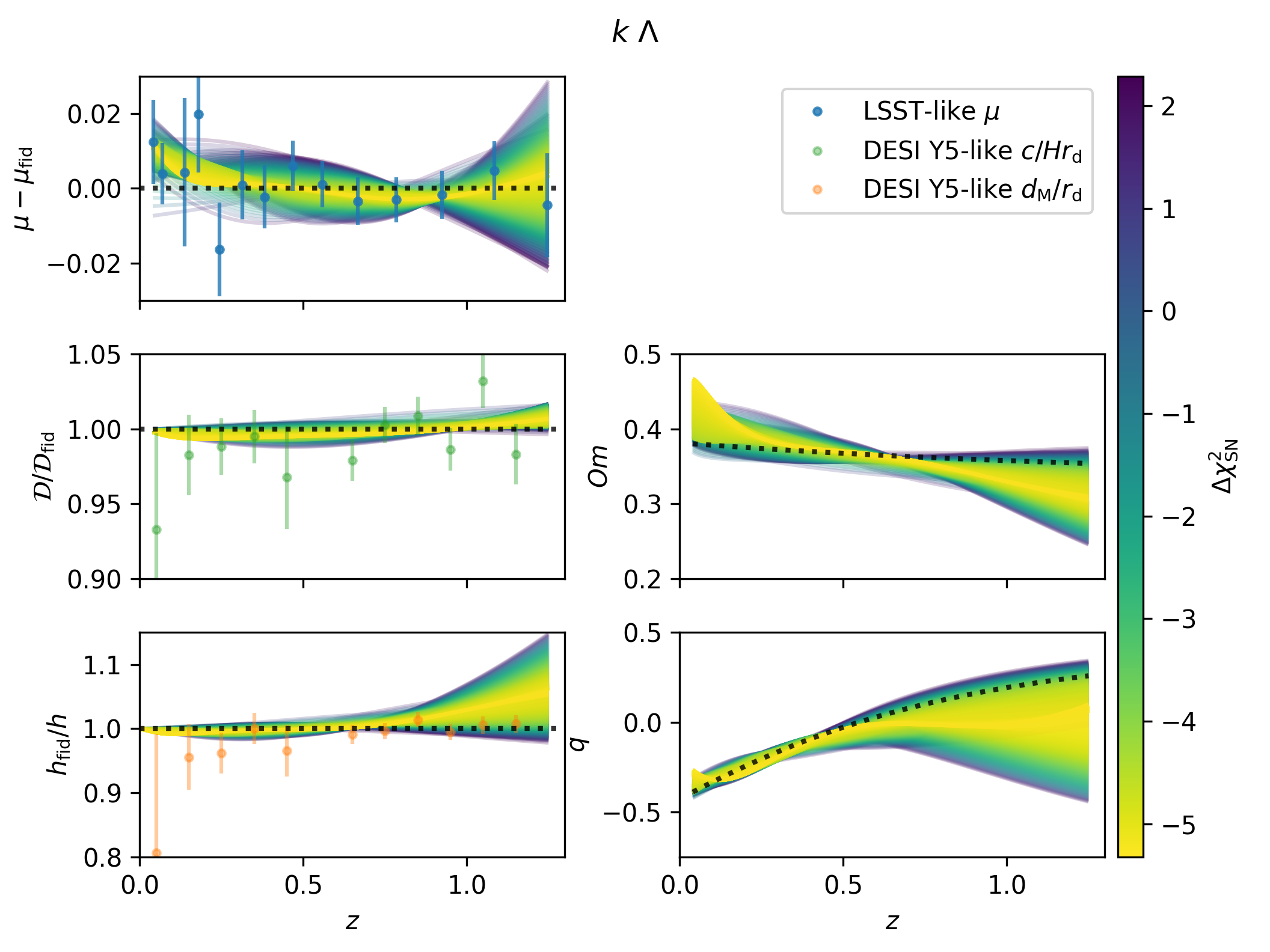}
    \caption{Same as Fig.~\ref{fig:smoothing_lcdm}, fiducial model: $k$-\lcdm\ 
    Note that in the bottom-left panel, the BAO data are $h_\text{fid}/h$, but the smooth SNIa reconstructions show $\DD'/\DD'_\text{fid}$.
    \label{fig:smoothing_klcdm} }
\end{figure}

\begin{figure}
    \centering
    \includegraphics[width=\textwidth]{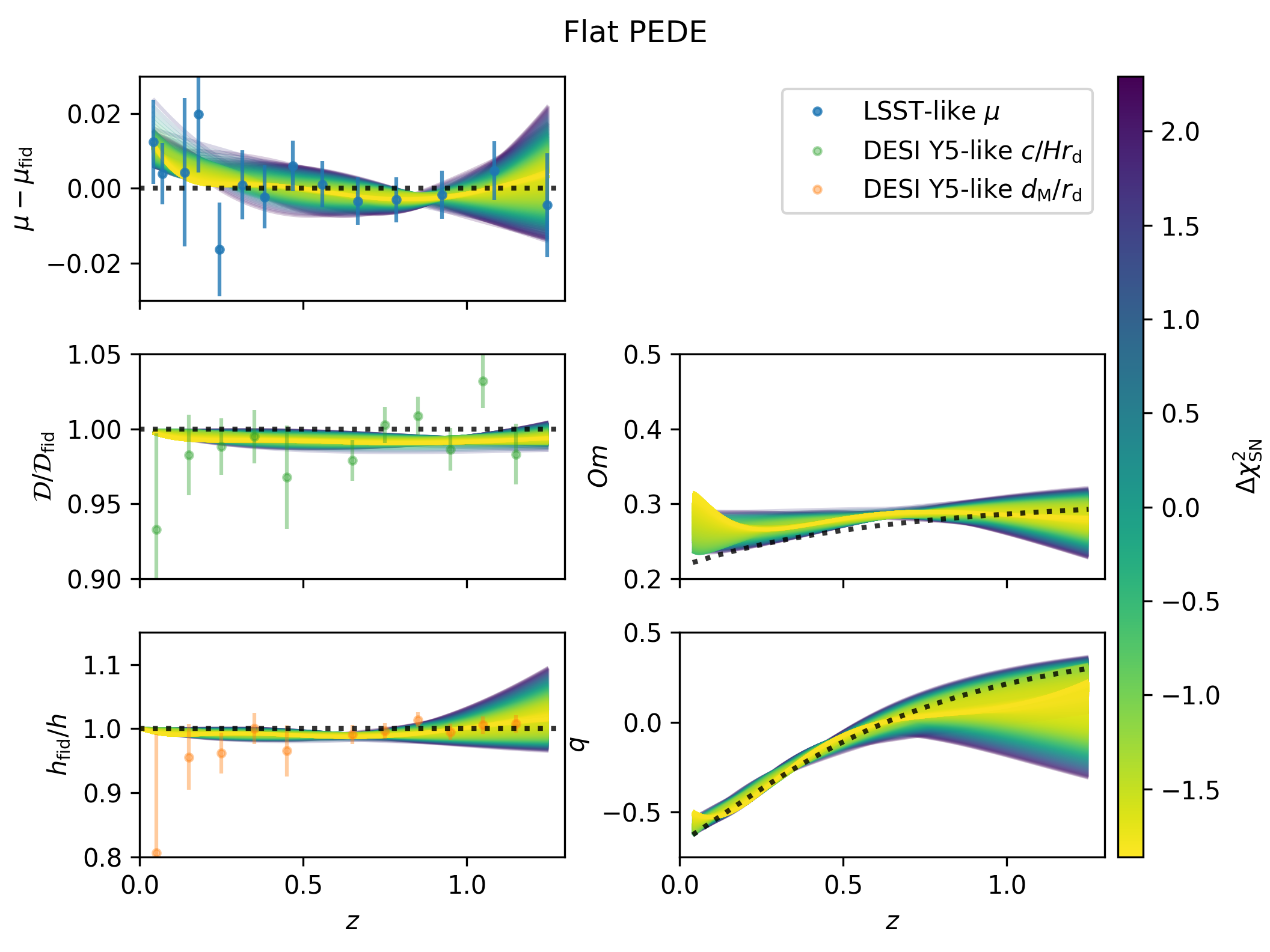}
    \caption{Same as Fig.~\ref{fig:smoothing_lcdm}, fiducial model: {Flat-}PEDE
    \label{fig:smoothing_pede} }
\end{figure}

\begin{figure}
    \centering
    \includegraphics[width=\textwidth]{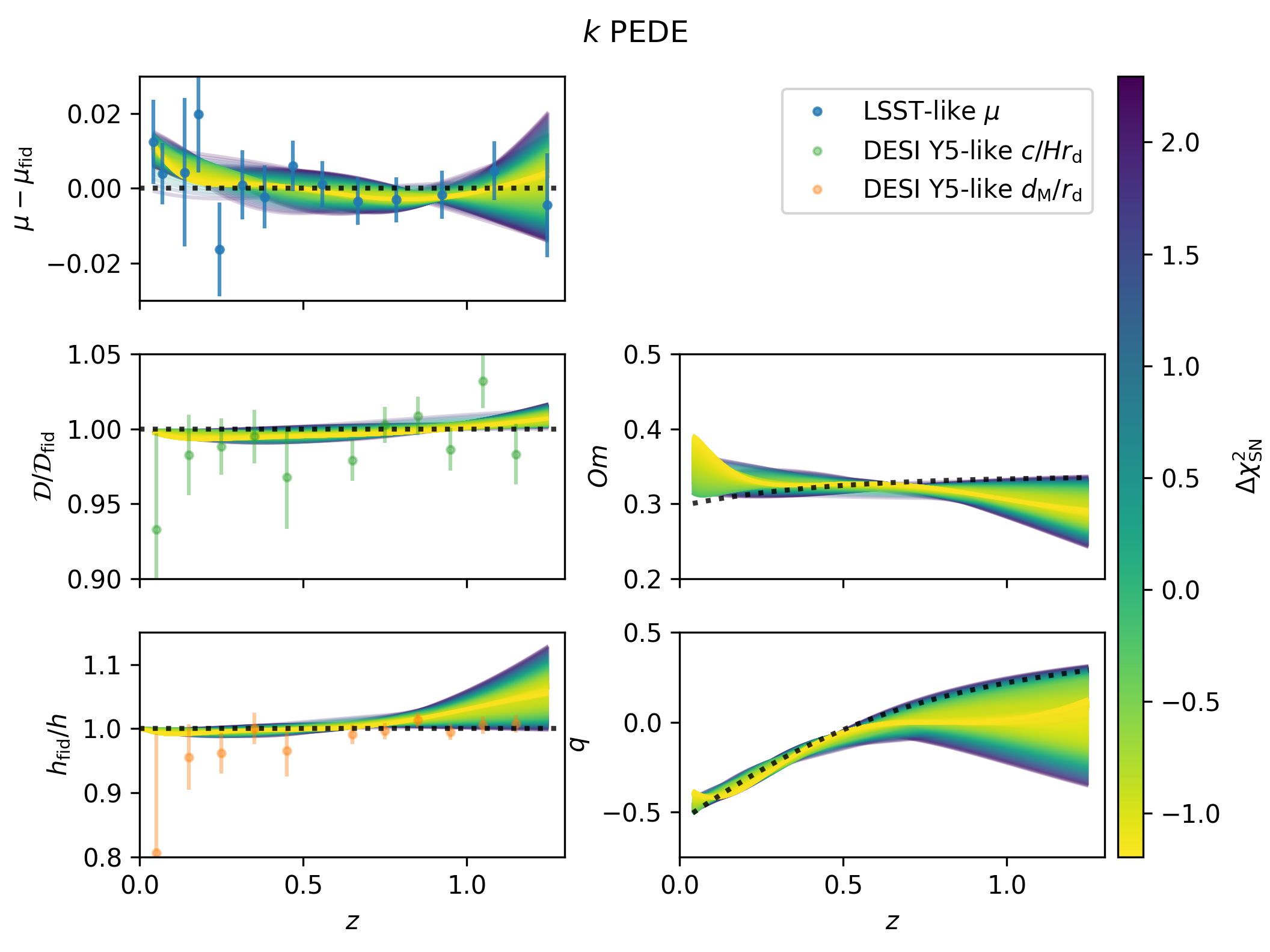}
    \caption{Same as Fig.~\ref{fig:smoothing_lcdm}, fiducial model: $k$-PEDE
    \label{fig:smoothing_kpede} }
\end{figure}

\subsection{Iterative Smoothing}
\label{sec:smooth}

We used the iterative smoothing method to reconstruct the distance-redshift relation from SNIa \cite{2006MNRAS.366.1081S,2007MNRAS.380.1573S,2018PhRvD..98h3526S}. 
The strength of this method is that we make no assumption regarding the evolution of dark energy.  

Starting with some initial guess $\hat\mu_0$, we iteratively calculate the reconstructed $\hat\mu_{n+1}$ at iteration $n+1$ with the inverse covariance matrix of the SNIa data $\tens{C}_\text{SN}^{-1}$:
\begin{subequations}\label{eq:smooth_SN}
\begin{align}
  \hat\mu_{n+1}(z) & = \hat\mu_n(z) + \frac{\vect{\delta\mu}_n^\intercal \cdot  \tens{C}_\text{SN}^{-1} \cdot \vect{W}(z)}
  {\one^\intercal \cdot \tens{C}_\text{SN}^{-1} \cdot \vect{W}(z)}, \label{eq:smooth}
  \intertext{where }
  \vect{W}_i(z) & =  {\exp{\left(- \frac{\ln^2\left(\frac
      {1+z}{1+z_i}\right)}{2\Delta^2}\right)}}
  \intertext{
 {is the smoothing kernel, }}
  \one & = (1,\dots,1)^\intercal\\
  \intertext{is a column vector,}
  \vect{\delta\mu}_n|_i & = \mu_i-\hat\mu_n(z_i).
\end{align}
\end{subequations}
The smoothing width is set to $\Delta = 0.3$ following previous analyses in \cite{2006MNRAS.366.1081S, 2007MNRAS.380.1573S, 2017JCAP...01..015L, 2018MNRAS.476.3263L, 2018PhRvD..98h3526S, 2020ApJ...899....9K, 2021JCAP...03..034K, 2022JCAP...03..047K}.
$W(z)$ can be understood as a Gaussian smoothing kernel of the variable $u=\ln(1+z)$, and the denominator as a normalization factor.

We define the $\chi^2$ value of the reconstruction $\hat{\mu}_n(z)$ at iteration $n$ as
\begin{equation}\label{eqn:chi2}
\chi^2_{\text{SN},n} = \vect{\delta \mu}_n^\intercal \cdot \tens{C}_\text{SN}^{-1} \cdot \vect{\delta \mu}_n.
\end{equation}
 
The results converge towards the solution preferred by the data, independently from the choice of the initial condition \cite{2006MNRAS.366.1081S,2007MNRAS.380.1573S,2017JCAP...01..015L,2018PhRvD..98h3526S}. 
Therefore, starting from different initial conditions, the results of the iterative smoothing method will approach this solution by different paths. 
More precisely, we start from the best-fit $k$-\lcdm, flat \lcdm, and flat \lcdm\ with fixed $\Omo=\{0,0.1,\dots,1\}$.
We collect all reconstructions with  $\Delta\chi^2_{\text{SN},n} =\chi^2_{\text{SN},n}-\chi^2_\text{SN,LCDM}< 2.3$ relative to the best-fit \lcdm. 
These reconstructions  then make up a non-exhaustive set of plausible distances and expansion histories. 

We also obtained the smoothed first and second derivatives of $\hat\mu$ as described in Appendix~\ref{sec:smoothder}.
The initial guess for the derivative of the distance modulus is
\begin{align}
    \deriv{}{\ln(1+z)}\hat{\mu}(z) & = \frac 5 {\ln 10} \left[1 + \frac{1+z}{h\DD} \sqrt{1+\Oko\DD^2}\right],\\
    \deriv{^2}{\ln(1+z)^2}\hat{\mu}(z) & = 
    \frac5{\ln 10} (1+z) \left[\frac{\DD'}\DD \left(1-(1+z)\frac{\DD'}\DD\right) + (1+z)\frac{\DD''}\DD\right]
    ,
\end{align}
where $'$ denotes $\diff / \diff z$.
The derivatives of $\mu$ are directly obtained during the smoothing process of the supernovae. 
The formalism of the smooth derivative is described in the appendix.
 
From $\hat\mu$, $\hat\mu'$, and $\hat\mu''$, we can  reconstruct 
\begin{align}
    \DD(z) & = \frac{H_0}c \frac{10^{\tfrac \mu 5-5}}{1+z}\\
    \DD'(z) &= \left[\frac {\ln 10}{5} \frac {\diff \mu}{\diff \ln(1+z)} -1\right] \frac {\DD(z)}{1+z}\mbox{, and}\\
    \DD''(z) & = -\frac {\DD'}{1+z} + \left[\frac{\ln 10}5 \deriv{^2\mu}{(\ln(1+z))^2}\DD  
    \right]
    +  \left[\frac{\ln 10}5 \deriv{\mu}{\ln(1+z)}-1\right] \frac{\DD'}{1+z}
\end{align}
In practice, $\hat\mu$ is reconstructed to an additive constant, and one has to normalize $\DD$, $\DD'$, and $\DD''$ so that $\DD'(z=0) = 1$.

From these relations, we can obtain $h$, $h'$, $\DD'$ via
\begin{align}
    \DD' &= \frac\DD{1+z}\left(\frac {\ln 10 }{5}\deriv \mu {\ln (1+z)} -1\right) \\
    h& = \frac{\sqrt{1+\Oko\DD^2}}{\DD'} \label{eq:hDprime} \\
    \intertext{and}
    h' & = \frac{h\DD}{\DD'} \left[\frac{\Oko}{h^2}-\frac{\DD''}{\DD}\right]. 
\end{align}
It is worth noting that, while we can obtain $\DD'$ and $\DD''$ from the smoothing of the supernovae directly, to obtain $h$ and $h'$, one needs to know $\Oko$. 
Therefore, wrongly assuming flatness will imply an error when inferring $h(z)$,  $h'(z)$, and  $q(z)$. 
This is just a restatement of the $\Ok$ test. 
{In addition, 
it is worth noticing that, while BAO provide both separate (but correlated) measurement of $\dcom/\rd$ and $\dhub/\rd$, $\DD$, $\DD'$ and $\DD''$ are obtained from the same measurements.}

\begin{figure}
        \centering
        \includegraphics[width=\textwidth]{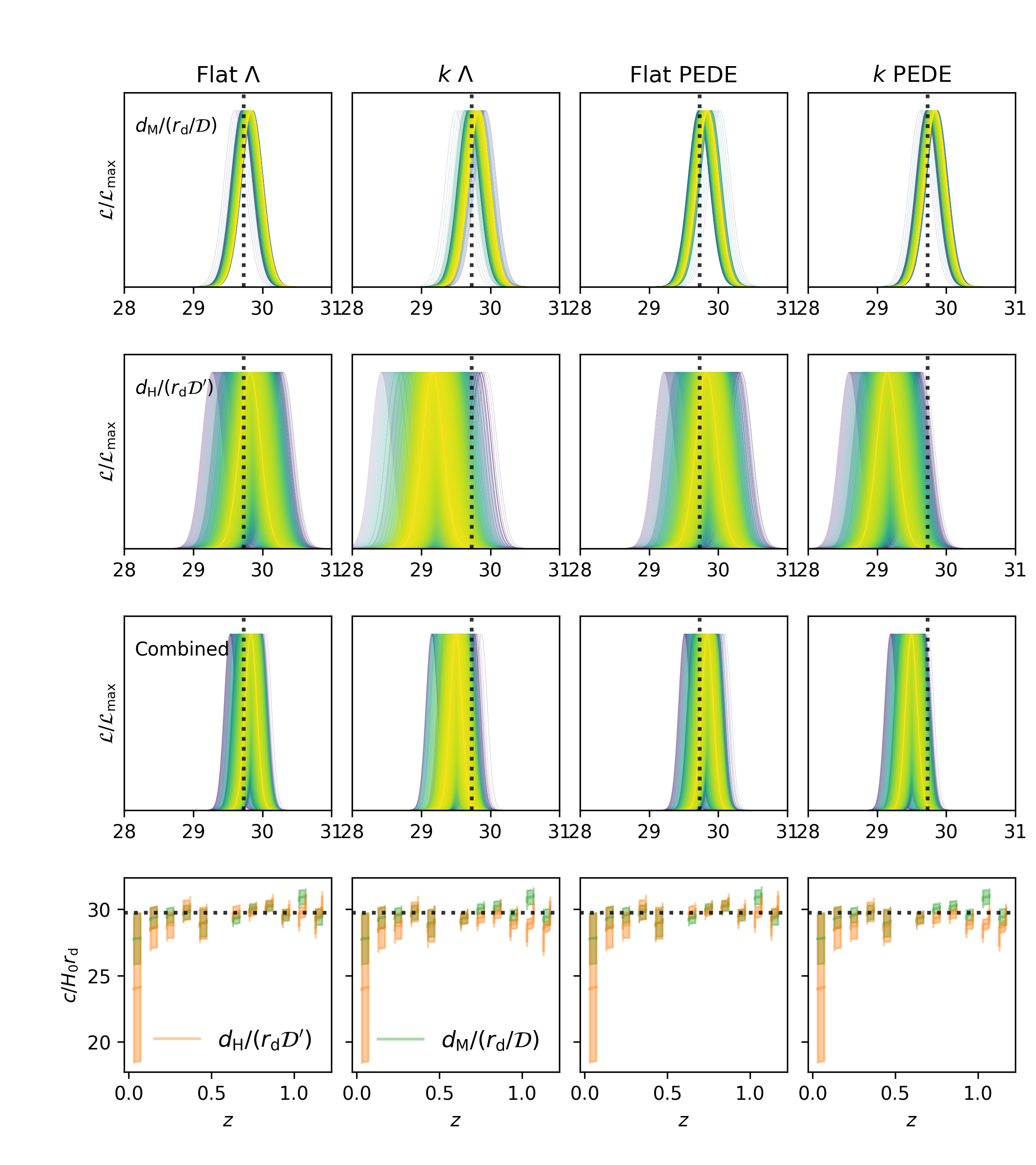}
        \caption{ 
        The bottom row shows our estimations of $c/\Hord$ for our two methods (Eqs.~\eqref{eq:HordD} and \eqref{eq:HordH}) at the BAO data points for the four fiducial cosmologies.
        The black dotted line is the fiducial value.
        The central line is the collection of central values of $c/\Hord$ for each reconstruction, and the coloured band is the one-sigma error from the BAO around the central value.
        The top rows show the likelihood  for estimate~\eqref{eq:HordD} at fixed iteration, and the second row for estimate~\eqref{eq:HordH}, and the third row shows the combination of both, taking into account the correlation.
        The colour bars of the top three rows are the same as that of the corresponding smooth figures~(\ref{fig:smoothing_lcdm} to \ref{fig:smoothing_kpede}). 
        \label{fig:H0rd}}
\end{figure}

\begin{figure}
    \centering
    \includegraphics[width=\textwidth]{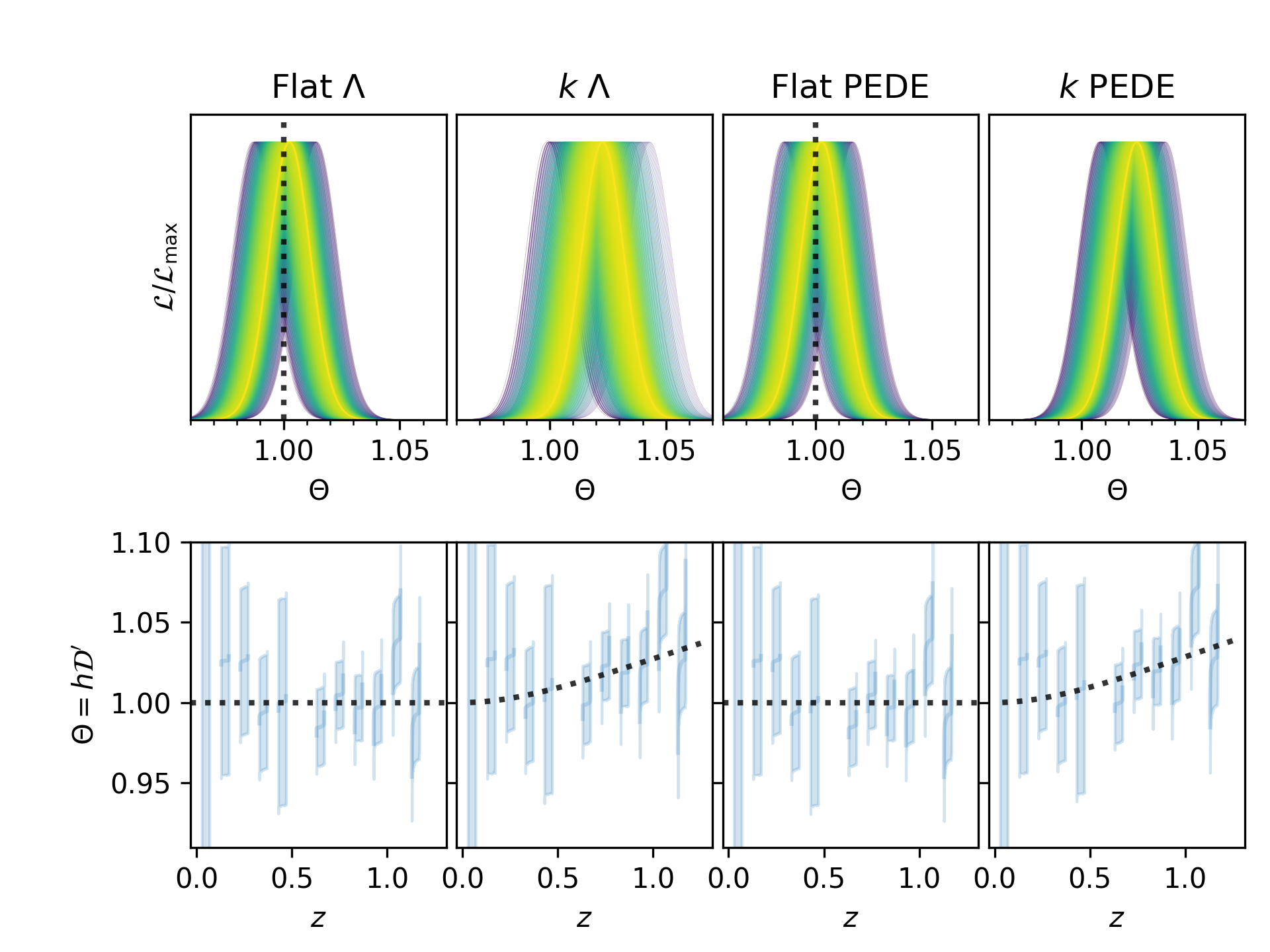}
    \caption{Curvature parameter: $\Theta$. 
    The bottom row shows $\Theta=h\DD'$ for our four fiducial cosmologies. 
    Like in Fig.~\ref{fig:H0rd}, the central value is obtained by combining the BAO with each iteration of the smooth SN reconstructions, and the error-band from the $1\sigma$ errors from the BAO.
    The top row shows the likelihood of $\Theta$ for each smooth reconstruction. 
    We note that in the second and fourth panels of the top row, there is no true value for $\Theta$ since $\Theta(z)=\sqrt{1+\Oko\DD^2}$ varies with redshift.}
    \label{fig:Theta}
\end{figure}

\begin{figure}
    \centering
    \includegraphics[width=\textwidth]{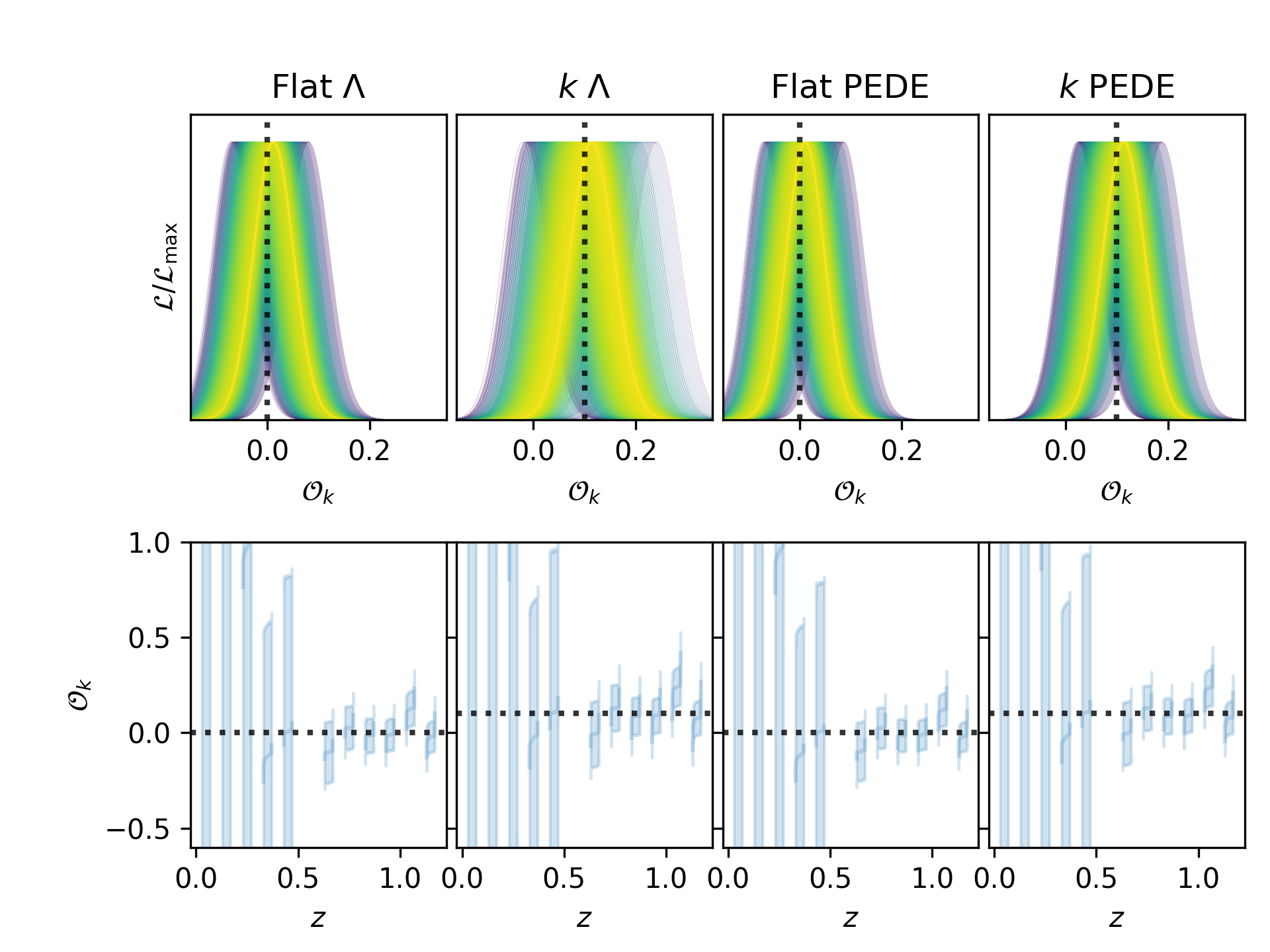}
    \caption{Curvature test: the \Ok\ diagnostic. Same legend as Figs.~\ref{fig:Theta}. The black dotted lines in the bottom row show the fiducial values of $\Oko$ in all three panels, $\Oko=0$ for the left and right, and $\Oko=0.1$ for the center.
    It is worth noting that in the $\Oko=0.1$ case, the \Ok\ diagnostic recovers the non-zero fiducial value of $\Oko$.}
    \label{fig:Ok}
\end{figure}

\section{Results}

\subsection{Supernova Smoothing}
Fig.~\ref{fig:smoothing_lcdm} shows the results of the smoothing for the flat-\lcdm\ fiducial model. 
Each line is an iteration of the smoothing color-coded by its $\Delta \chi^2$ with respect to the best-fit \lcdm\ model.  
The top-left panel shows the reconstructed distance moduli with the mock SNIa data. 
The middle- and bottom-left panels show the comoving distance $\DD(z)/\DD_\text{fid}(z)$ and the inverse of the expansion history $h_\text{fid}(z)/h(z)$ normalized to the fiducial values.
Again, the colored lines come from the smoothing of the SNIa, while the data points with error-bars come from the BAO. 
The middle- and bottom-right panels show $\Om$ and the deceleration parameter $q$. 
We are able to reconstruct the fiducial quantities. 
We report the $\Delta\chi^2_\text{fid}=\chi^2_\text{fid}-\chi^2_{\lcdm}$ in the last column of Table~\ref{tab:fidcosmo}.

Figs.~\ref{fig:smoothing_klcdm},~\ref{fig:smoothing_pede}, \ref{fig:smoothing_kpede} show the same results for the $k$\lcdm, flat-PEDE, and $k$-PEDE  fiducial models. 
In all cases, we are able to reconstruct the $\Om$ and $q$ parameters. 
However, it is interesting to look closely at the two non-flat cases. 
Since the assumption of flatness that goes into eq.~\eqref{eq:Dflat} is not true in this case, calculating $\DD'$ does not provide $h$ (see eq.~\eqref{eq:hDprime}). 
This in turn will bias our inference of $\Om$ and $q$ as well as the curvature test. 
At high-enough redshifts $z\geq 1$, the deviation between $\DD'$ and $1/h$ (i.e., $\sqrt{1+\Oko\DD^2}$) becomes important. This deviation between the $h_{\rm fid}/h$ data from the BAO and the $\DD'/\DD$ reconstructions from the SNIa are the litmus test for flatness, or signal for curvature that we are looking for.
Interestingly, the reconstructions are still consistent with the fiducial $q$. 
 
\subsection{\Hord}
 
The bottom row of Fig. \ref{fig:H0rd} shows our two estimations of $c/\Hord$ from eqs~\eqref{eq:HordD} and ~\eqref{eq:HordH} for each fiducial model. 
At each BAO data point, we plot the central value and the $\pm 1\sigma$  band for each reconstruction $n$ with $\Delta\chi_{\text{SN},n}^2=\chi_{\text{SN},n}^2-\chi^2_\text{SN, \lcdm}<2.3$.  
It is worth stressing here that at every BAO point, we have $N_\text{rec}$ estimations of $c/\Hord$, corresponding to $N_\text{rec}$ reconstructions of the distance-redshift relations from the SNIa (combined with the mock BAO data point). 
Each of these estimates also has an associated error bar, shown as an error band on the figure. 
At each redshift point, the reconstructed $c/\Hord$ are re-sorted by increasing central value for aesthetic reasons. 
The fiducial value $c/\Hord^\text{fid}=29.73$ is shown as a dashed line. 
For the four fiducial cosmologies, the estimations from eq.~\eqref{eq:HordD} are consistent with $\Hord^\text{fid}$.

Rows 1 to 3 show the (normalized) likelihood for iteration $n$
\begin{subequations}
    \begin{align}
        \mathcal L_{n,i}(x) & \propto \exp\left(- \frac{\chi^2_n}2\right)\mbox{, where} \\
        \chi^2_n & = - \frac 1 2 (x\one- \vect X_i)^\intercal \cdot \tens{C}_i^{-1} \cdot (x\one- \vect X_i), \\
        \vect X_1^\intercal &= 
        \begin{pmatrix} \frac 1 {\DD(z_1)} \frac {\dcom(z_1)}{\rd}, \dots,  \frac1{\DD(z_N)} \frac {\dcom(z_N)}{\rd} \end{pmatrix},\label{eq:x1}\\
        \vect X_2^\intercal & = \begin{pmatrix}  \frac 1 {\DD'(z_1)} \frac{\dhub(z_1)}{\rd} , \dots,  \frac 1 {\DD'(z_N)} \frac {\dhub(z_N)}{\rd} \end{pmatrix}, \label{eq:x2}\\
        \vect X_3 & = \begin{pmatrix}
            \vect X_1\\
            \vect X_2
        \end{pmatrix}\label{eq:x3}
    \end{align}
\end{subequations}
are the data vectors, and $N=11$ is the number of BAO data points.
The error propagation, including the covariance between the two BAO modes, is detailed in Appendix~\ref{sec:errorprop}. 
In the case of eq.~\eqref{eq:HordD} (row 1), the three fiducial cosmologies yield similar reconstructions, regardless of the curvature and DE.  
This is to be expected, since both distances from the SNIa and BAO should agree regardless of the curvature.
It is worth paying attention to the $k$-$\Lambda$ and $k$-PEDE cases (second and fourth columns). 
While estimator~\eqref{eq:HordD} (row 1) is totally consistent with the fiducial value, eq.~\eqref{eq:HordH} (row 2) is centred around 28 to 29, and most of the reconstructions are over $2\sigma$ away from the fiducial value, showing an inconsistency between the two estimators. 
This comes from the fact that while the radial mode of the BAO provides us with $1/h$, the SNIa   provide us with $\DD'$, which do not coincide in a non-flat FLRW universe, as seen above. 
Indeed, we can see in the bottom row that the values of $ \dhub/(\rd \DD')$ diverge from the fiducial values at high redshift, reflecting equation~\eqref{eq:hDprime}. 
For each quantity $A$, we report the minimal and maximal values of the precision $\sigma_A$ and the spread $\Delta A= (A-A_\text{fid})$ taken over all reconstructions in Table~\ref{tab:Prec}. 
Namely, for a given reconstruction $n$, we obtain the likelihood in the top three panels, and obtain the corresponding precision and spread. 
We then report the minimal and maximal value over the whole reconstructions.
In the case of $X_i$ (eqs.~\eqref{eq:x1} to \eqref{eq:x3}, we normalize the result by the fiducial value. 
The third row shows a greater precision by combining the two estimators despite the correlation between the radial and transverse BAO modes.

\begin{table}[]
    \centering
    \begin{tabular}{ccccc}
    \toprule
     Quantity & 
     flat-$\Lambda$ & $k$-$\Lambda$ & 
     PEDE & 
     $k$-PEDE\\
    \midrule 
$\sigma_{X_1}/X_\text{fid}$ & 
0.52\% & 
0.52\% & 
 0.52 -- 0.53\% & 
 0.52\% \\
$\Delta {X_1}/X_\mathrm{fid}$ &
-0.44\% -- 0.44\% &
-0.77\% -- 0.68\% & 
-0.05\% -- 0.98\% &  
-0.37\% -- 0.49\% \\ 
\midrule
$\sigma_{X_2}/X_\text{fid}$ & 
0.50 -- 0.52\% & 
0.49 -- 0.51\%  & 
  0.50 -- 0.52\%  & 
  0.49 -- 0.51\%  \\
$\Delta {X_2}/X_\mathrm{fid}$ &
-1.55\% -- 1.94\% &
 -4.40\% -- 0.72\%  &
 -1.74\% -- 2.04\% &  
-3.82\% -- -0.18\%   \\  
\midrule
$\sigma_{X_3}/X_\mathrm{fid}$ & 
 0.28 -- 0.29 \%         & 
 0.28\%         & 
0.28 -- 0.29\%         & 
0.28\% \\
  $\Delta {X_3}/X_\mathrm{fid}$ & 
-0.68\% -- 1.06\% & 
-1.96\% -- 0.47\% & 
-0.75\% -- 1.24\% & %
-1.82\% -- 0.02\%\\ %
\midrule
$\sigma_{\Ok}$ & 
0.037 -- 0.040 & 
0.040 -- 0.045 & 
0.035 -- 0.038 & 
0.039 -- 0.042 \\ 
$\Delta {\Ok}$  & 
   0.004 -- 0.015  & 
   -0.095 -- -0.031&          
    0.004 -- 0.016 &
 -0.095 -- -0.055   \\
\midrule
$\sigma_\Theta$ & 
0.009 & 
0.009  & 
0.009  & 
0.009  \\ 
  $\Delta \Theta$ &
   -0.026 -- 0.029 &
  --- &
   -0.0277 -- 0.0332&
  --- \\
\bottomrule 
    \end{tabular}
    \caption{Table of precision and spread for our four fiducial cosmologies.
    The reported values are the minimum and the maximum values of the $1\sigma$ CL of the likelihoods in Figs.~\ref{fig:H0rd} to~\ref{fig:Ok} over all reconstructions. 
    The spread of quantity $A$, noted as $\Delta A$ is defined as the difference between the peak of the likelihood and  the fiducial value. 
    For simplicity, we note $X=c/\Hord$, and the indices refer to the different estimators (see eqs.~\eqref{eq:x1} to \eqref{eq:x3}).
    The precisions for our three estimates of $X=c/\Hord$ are relative to the true value, while the precisions and spreads in $\Ok$ and $\Theta$ are absolute.
    We note that in the non-flat cases ($k$-$\Lambda$ and $k$-PEDE), there is no \emph{fiducial} value for $\Theta$ since $\Theta$ evolves with redshift, so we do not report the spread.
    }
    \label{tab:Prec}
\end{table}

\subsection{Curvature Test}

The bottom panel of Fig.~\ref{fig:Theta} shows our estimate of $\Theta(z) = h\DD' = \sqrt{1+\Oko\DD^2}$ for the three cosmologies. 
The top panel shows the normalized likelihood of $\Theta(z)$ over $z$ for a given iteration of the smoothing, colour-coded by $\Delta\chi^2$ of the SNIa reconstruction. 
As expected, for the Flat-$\Lambda$ (left) and PEDE (right) cases, our estimates are consistent with $\Theta=1$. 
The interesting cases are the $k$-$\Lambda$ and $k$-PEDE cases (second and fourth columns).
We can see in the bottom panel a trend of $\Theta(z)$ deviating from 1 as $z$ increases. 
Similarly, in the top panel, the likelihoods peak at higher values than in the other two cases, and only marginally consistent with $\Theta=1$.  

Fig.~\ref{fig:Ok} shows similar plots for the \Ok\ parameter. 
Similarly to $\Theta$, the Flat-$\Lambda$
 and PEDE case are perfectly consistent with flatness ($\Oko=0$), while the $\Oko=0.1$ case is only marginally consistent with $\Oko=0$ and perfectly consistent with $\Oko=0.1$, showing the constraining power of this test. 

In the bottom panels of both figures, at $z\gtrsim 0.5$, the error bars are very small and the reconstructions are perfectly consistent with the fiducial cosmology in all three cases. 
At low redshift, the large error bars do not constrain $\Theta$ or $\Oko$.

In order to assess the power and significance of the test, we perform a hypothesis test. 
One must bear in mind that the problem is not symmetric here: the null hypothesis $H_0$ to test is ``$\Omega_{k0} = 0$''.
For the two flat models, we are interested in type I errors: what is $p_I$, the probability of wrongly rejecting flatness? For the two non-flat models, the question is ``What is the probability of failing to reject flatness?'', i.e., type II errors, with probability $p_{II}$. 

To assess these, we resimulate 1000 realizations of the data. 
Since different iterations of the smoothing are correlated, one cannot simply combine them. 
Instead, we choose for each realization the one with the lowest $\Delta\chi^2$. We then calculate, for each realization, the likelihood of $\Oko \pm \sigma_{\Oko}$. 
For a given value of {the significance} $\alpha$, we want to measure $p_I$ and $p_{II}$, assuming a Gaussian likelihood for $\Oko$, i.e., $p_I = 2P(X > |\Oko|)$, and $p_{II} = 1-2P(X>|\Oko|)$. 
The results are reported on Table~\ref{tab:Prec}.
While the probabilities of type I errors are reasonable, for a given $\alpha$, the probability of type II errors are typically 40\% ($\alpha=0.05$) to 60\% ($\alpha=0.01$). 
While a LSST 3 year+DESI 5 years survey is not sufficient to rule out flatness at a high confidence level for curvature values of 0.1, it is important to keep in mind that this is only using low-redshift data, and does not need any assumption regarding the dark energy model.

\begin{table}[]
    \centering
    \begin{tabular}{ccccc}
    \toprule
        &  $\Lambda$  &  $k$-$\Lambda$ & PEDE &   $k$-PEDE\\
     $\alpha$   & $p_{I}$ & $p_{II}$ & $p_{I}$ & $p_{II}$\\
    \midrule
        0.05 & 0.068 & 0.363 & 0.065 & 0.334 \\
        \midrule
        0.01 & 0.016 & 0.595 & 0.014 & 0.548\\
    \bottomrule
    \end{tabular}
    \caption{Probabilities of type I and type II errors for the four fiducial models.  }
    \label{tab:pval}
\end{table}

\section{Summary and Conclusion}

We combined forecast BAO data for a DESI 5-year-like survey and a realistic forecast for LSST SNIa to assess the ability of the joint surveys to provide model-independent litmus tests of the flat \lcdm\ model. 
For this purpose, we generated mock data for  four cosmologies: flat \lcdm\, a non-flat \lcdm\ universe with $\Oko=0.1$, and two examples of dynamical dark energy, the PEDE model, with $\Oko=0$ and $\Oko=0.1$.

We applied the iterative smoothing method to reconstruct the expansion history in a model-independent manner, only assuming flatness. 

The expected quality of LSST SNIa will allow to reconstruct very accurately certain diagnostics such as $\Om$ and $q$. 
In addition, we quantify the ability of the joint surveys to constrain $c/\Hord$. 
We then proceed to testing the FLRW metric, measuring $\Theta$ and $\Ok$. 
The combination of LSST 3-year and DESI 5-years will allow to constrain $O_k$ to within $\sim 0.035$, and $\Theta$ to within 0.0075, with a spread due to the SNIa uncertainties of up to $\pm 0.1$ and 0.03 respectively. 
It is worth noting that the credible interval {on $\Oko$} reported by DESI DR1 is  $(-0.078,0.068)$ for the flat \lcdm\ model without any external data. 
We show that with the expected five-year data accuracy, a model-independent precision on the curvature parameter will be competitive.

In the case of the $\Oko=0.1$ universes, the flatness assumption of eq.~\eqref{eq:Dflat} is now incorrect, which leads to a biased expansion history and inconsistent estimations of $c/\Hord$ from our two methods. 
Therefore, $\Hord$ itself can be seen as a curvature test.
An inconsistency between the two estimates of $\Hord$, or with $\Ok(z) = $ constant or $\Theta(z) = 1$ can just be interpreted either as evidence of departure from flatness, or inconsistency between the two data sets.

\appendix

\section{Smooth Derivatives}\label{sec:smoothder}

An interesting property of smoothing is that the derivative of the smooth function can be readily obtained from the derivative of the smoothing kernel.

Suppose we observe a function $y_i = f(t) + \vect{\epsilon_i}$, where $\vect\epsilon\sim\mathcal N(0,\mat{C})$. 
We want to reconstruct the smooth function $\hat f(t)$ and its successive derivative $\hat f^{(i)}(t)$.

Since smoothing is essentially a convolution by a smoothing kernel, one can straightforwardly define the smooth reconstruction of the successive derivatives as
\begin{align}
    \hat y_{n+1}(t) & = \hat y_n(t) + \frac {B(t)}{A(t)}, 
    \intertext{where}
    B & = \vect{\delta y}_n\cdot \mat C^{-1}\cdot \vect{W(t)}, \text{and}\\
    A & = \one^\intercal\cdot \mat C^{-1}\cdot \vect{W(t)}.
    \intertext{The first and second smooth derivatives is obtained by deriving the previous equation with respect to $t$:}
    \widehat{y}'_{n+1}(t) & = \widehat{y'}_n(t) + \frac {B'}{A}  - \frac{BA'}{A^2}, \\
    \widehat{y}''_{n+1}(t) & = \widehat{y''_n}(t) + \frac{B''}{A} - \frac{2A'B'+A''B}{A^2}+ \frac{2A'^2B}{A^3},
    \intertext{where}
    B' & = \vect{\delta y}_n^\intercal\cdot \mat C^{-1} \cdot \vect W'(t), \\
    A' & = \one^\intercal\cdot \mat C^{-1} \cdot \vect W'(t),\\
    B''&  = \vect{\delta y}_n^\intercal\cdot \mat C^{-1} \cdot \vect W''(t), \text{and} \\
    A'' & = \one^\intercal\cdot \mat C^{-1} \cdot \vect W''(t).
\end{align}

In this work, we apply the algorithm to the LSST simulated SNIa with $t=\ln (1+z)$.

\section{Error propagation}
\label{sec:errorprop}

The errors are propagated using the usual formula: for $\vect Y=\vect f(\vect X)$, 
the covariance $\mat {C}_{Y}$ is given by 
\begin{align}
    \mat C_ {Y} & = \mat  J^\intercal \cdot  \mat {C}_ {X} \cdot \mat J, 
\end{align}
where \mat{J} is the Jacobian of $\vect f${, and \vect{Y} stands for c/\Hord, $\Theta$, or $\Ok$.}

Since the correlation between two redshift bins in the BAO data is negligible, we can propagate the errors independently at fixed redshift $z$. We then have
\begin{align}
\mat{C}_{c/\Hord} & = \mat J_{c/\Hord}^\intercal \cdot \mat{C}_\mathrm{BAO} \cdot \mat J_{c/\Hord}\\ 
\intertext{where}
\mat J_{c/\Hord}& = \begin{pmatrix}
    \frac1{\DD(z)}  & 0 \\
    0 & \frac1{\DD'(z)}
\end{pmatrix}.
\end{align}

For $\Theta$ and $\Ok$, the covariance matrix is simply the variance. The variance of $\Theta$ is obtained by
\begin{align}
    \sigma_\Theta^2 & = \mat J_\Theta^\intercal \cdot \mat C_\text{BAO} \cdot \mat J_\Theta, 
    \intertext{where}
    \mat J_\Theta & = \Theta
    \begin{pmatrix}
        \frac {\rd}{\dcom} \\
        \frac{\rd}{\dhub}
    \end{pmatrix},
\end{align}
and that of $\Ok$ by
\begin{align}
 \sigma_{\Ok} = 2 \frac{\Theta}{\DD^2} \sigma_\Theta.   
\end{align}

In all cases, the total $\chi^2$ for reconstruction $n$ is obtained by
\begin{align}
    \chi^2_n = \chi^2_{\text{SN},n} + \chi^2_{\text{BAO},n}. 
\end{align}

\acknowledgments

The authors thank Kushal Loda for constructive discussions.
B.~L. acknowledges the support of the National Research Foundation of Korea (NRF-2022R1F1A1076338 and RS-2023-00259422). 
A.M. thanks  Richard Kessler and the LSST DESC TD team for the production of the LSST dataset. A. S. would like to acknowledge the support by National Research Foundation of Korea NRF2021M3F7A1082056 and the support of the Korea Institute for Advanced Study (KIAS) grant funded by the
government of Korea.

\bibliographystyle{JHEP}
\bibliography{biblio}

\end{document}